\documentclass[journal,transmag,twoside]{IEEEtran}
\usepackage[T1]{fontenc}
\usepackage{times}
\usepackage{amsfonts,amsmath,amssymb,amsbsy}
\usepackage{graphicx,color,array,cite}
\usepackage[colorlinks=true, linkcolor=blue, citecolor=red, urlcolor=magenta]{hyperref}
\let\mathbf=\boldsymbol

\def\emph#1{\textcolor{red}{#1}}
\def\emph#1{\textcolor{black}{#1}}
\begin{document}

\title{Magnetic Skyrmion Transport in a Nanotrack With Spatially Varying Damping and Non-adiabatic Torque}

\author{\IEEEauthorblockN{%
Xichao Zhang\IEEEauthorrefmark{1,2},
Jing Xia\IEEEauthorrefmark{1},
G. P. Zhao\IEEEauthorrefmark{3},
Xiaoxi Liu\IEEEauthorrefmark{4},
and
Yan Zhou\IEEEauthorrefmark{1}}
\IEEEauthorblockA{\IEEEauthorrefmark{1}School of Science and Engineering, The Chinese University of Hong Kong, Shenzhen 518172, China}
\IEEEauthorblockA{\IEEEauthorrefmark{2}School of Electronic Science and Engineering, Nanjing University, Nanjing 210093, China}
\IEEEauthorblockA{\IEEEauthorrefmark{3}College of Physics and Electronic Engineering, Sichuan Normal University, Chengdu 610068, China}
\IEEEauthorblockA{\IEEEauthorrefmark{4}Department of Information Engineering, Shinshu University, Wakasato 4-17-1, Nagano 380-8553, Japan}
\thanks{The first two authors contributed equally to this work. Corresponding authors: X. Liu (email:~liu@cs.shinshu-u.ac.jp) and Y. Zhou (email:~zhouyan@cuhk.edu.cn).}}


\IEEEtitleabstractindextext{%
\begin{abstract}
Reliable transport of magnetic skyrmions is required for any future skyrmion-based information processing devices. Here we present a micromagnetic study of the in-plane current-driven motion of a skyrmion in a ferromagnetic nanotrack with spatially sinusoidally varying Gilbert damping and/or non-adiabatic spin-transfer torque coefficients. It is found that the skyrmion moves in a sinusoidal pattern as a result of the spatially varying Gilbert damping and/or non-adiabatic spin-transfer torque in the nanotrack, which could prevent the destruction of the skyrmion caused by the skyrmion Hall effect. The results provide a guide for designing and developing the skyrmion transport channel in skyrmion-based spintronic applications.
\end{abstract}
}

\maketitle
\IEEEdisplaynontitleabstractindextext
\IEEEpeerreviewmaketitle

\section{Introduction}
\label{se:Introduction}

Magnetic skyrmions are quasiparticle-like domain-wall structures with typical sizes in the sub-micrometer regime~\cite{Braun_AiP2012,Nagaosa_NNANO2013,Liu_CPB2015,Wiesendanger_Review2016,Kang_Review2016,Finocchio_Review2016,Seki_BOOK2016}. They are theoretically predicted to exist in magnetic metals having antisymmetric exchange interactions~\cite{Roszler_NATURE2006}, and confirmed by experiments~\cite{Muhlbauer_SCIENCE2009,Yu_NATURE2010} just after the turn of the twenty-first century. Isolated skyrmions are expected to be used to encode information into bits~\cite{Fert_NNANO2013}, which might lead to the development of novel spintronic applications, such as the racetrack memories~\cite{Parkin_SCIENCE2008,Iwasaki_NNANO2013,Sampaio_NNANO2013,Xichao_SREP2015A,Tomasello_SREP2014,Xichao_NCOMMS2016,Du_NCOMMS2015,Muller_NJP2016}, storage devices~\cite{Beg_SREP2015,Yuan_SREP2016,Bazeia_JMMM2017}, and logic computing devices~\cite{Xichao_SREP2015B}.

The write-in and read-out processes of skyrmions in thin films are realizable and controllable at low temperatures~\cite{Romming_SCIENCE2013,Hanneken_NNANO2015,Crum_NCOMMS2015}. A recent experiment has realized the current-induced creation and motion of skyrmions in Ta/CoFeB/TaO trilayers at room temperature~\cite{Wanjun_SCIENCE2015}. Experimental investigations have also demonstrated the increased stability of skyrmions in multilayers~\cite{Woo_NMATER2016,Moreau-Luchaire_NNANO2016,Boulle_NNANO2016}, which makes skyrmions more applicable to practical room-temperature applications.

However, the skyrmion experiences the skyrmion Hall effect (SkHE)~\cite{Zang_PRL2011,Wanjun_ARXIV2016}, which drives it away from the longitudinal direction when it moves in a narrow nanotrack. As a consequence, in the high-speed operation, the transverse motion of a skyrmion may result in its destruction at the nanotrack edges~\cite{Xichao_NCOMMS2016,Purnama_SREP2015,Xichao_SREP2016,Tretiakov_PRL2016,Xichao_SyAFSk2016}. Theoretical and numerical works have proposed several intriguing methods to reduce or eliminate the detrimental transverse motion caused by the SkHE. For example, one could straightforwardly enhance the perpendicular magnetic anisotropy near the nanotrack edges to better confine the skyrmion motion~\cite{Purnama_SREP2015}. An alternative solution is to transport skyrmions on periodic substrates~\cite{Reichhardt_PRB2015A,Reichhardt_PRB2015B,Reichhardt_PRB2016,Reichhardt_NJP2015}, where the skyrmion trajectory can be effectively controlled. Moreover, by constructing antiferromagnetic skyrmions~\cite{Tretiakov_PRL2016,Xichao_SREP2016} and antiferromagnetically exchange-coupled bilayer skyrmions~\cite{Xichao_NCOMMS2016,Xichao_SyAFSk2016}, the SkHE can be completely suppressed. Recently, it is also found that the skyrmionium can perfectly move along the driving force direction due to its spin texture with a zero skyrmion number~\cite{Xichao_PRB2016,Komineas_PRB2015}.

In this paper, we propose and demonstrate that a skyrmion guide with spatially sinusoidally varying Gilbert damping and/or non-adiabatic spin-transfer torque (STT) coefficients can be designed for transporting skyrmions in a sinusoidal manner, which is inspired by a recent study on the magnetic vortex guide~\cite{Yuan_AIPADV2015}, where the vortex core motion is controlled via spatially varying Gilbert damping coefficient. The results provide a guide for designing and developing the skyrmion transport channel in future spintronic devices based on the manipulation of skyrmions.

\section{Methods}
\label{se:Methods}

\begin{figure}[t]
\centerline{\includegraphics[width=0.5\textwidth]{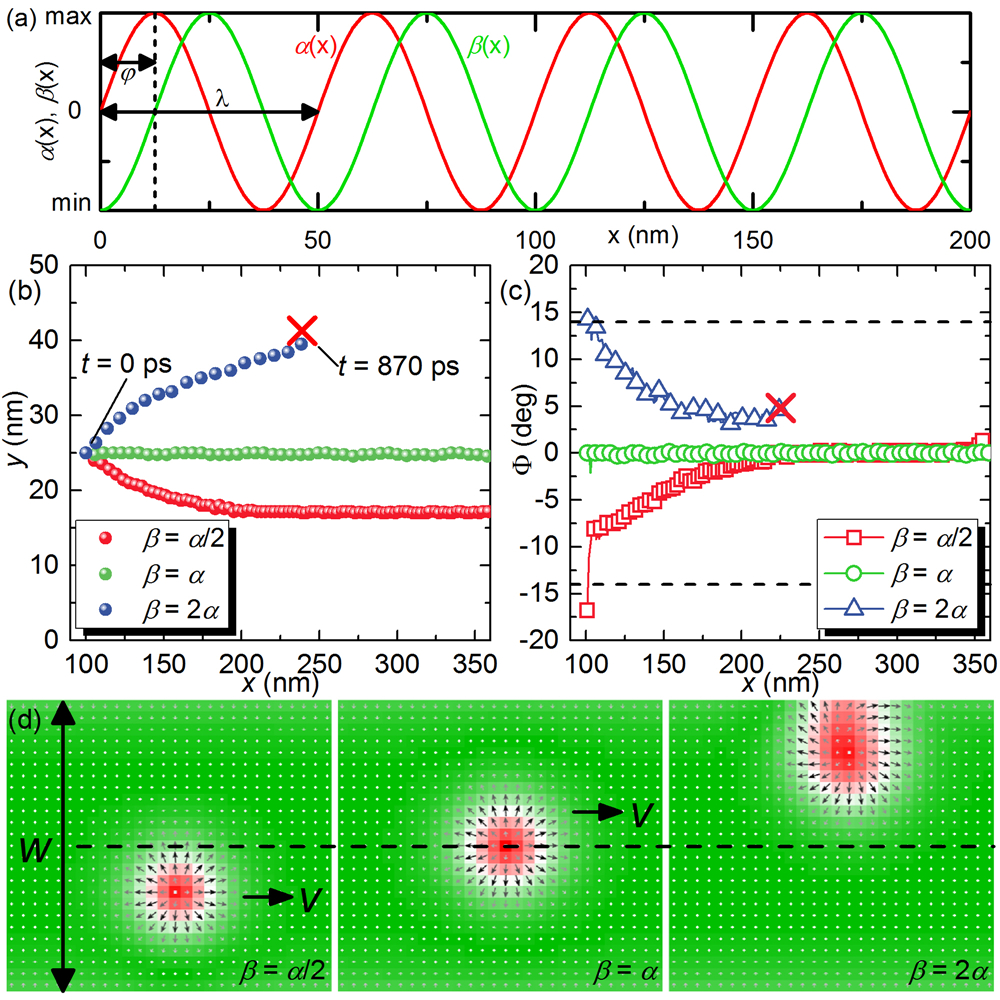}}
\caption{%
(a) The magnetic damping coefficient $\alpha(x)$ and non-adiabatic STT coefficient $\beta(x)$ as functions of $x$ in the nanotrack.
(b) Trajectories of current-driven skyrmions with $\beta=\alpha/2=0.15$, $\beta=\alpha=0.3$, and $\beta=2\alpha=0.6$. Dot denotes the skyrmion center. Red cross indicates the skyrmion destruction.
(c) Skyrmion Hall angle $\Phi$ as a function of $x$ for skyrmion motion with $\beta=\alpha/2=0.15$, $\beta=\alpha=0.3$, and $\beta=2\alpha=0.6$. The dashed lines indicate $\Phi=\pm 14^{\circ}$.
(e) Real-space top-views of skyrmion motion with $\beta=\alpha/2=0.15$, $\beta=\alpha=0.3$, and $\beta=2\alpha=0.6$. $w$ and $v$ denote the nanotrack width and velocity direction, respectively. The dashed line indicates the central line of the nanotrack. The skyrmion is destroyed at $t=870$ ps when $\beta=2\alpha=0.6$. The out-of-plane magnetization component is represented by the red ($-z$)-white ($0$)-green ($+z$) color scale.
}
\label{FIG1}
\end{figure}

Our simulation model is an ultra-thin ferromagnetic nanotrack with the length $l$ and the width $w$, where the thickness is fixed at $1$ nm. We perform the simulation using the standard micromagnetic simulator, i.e., the 1.2 alpha 5 release of the Object Oriented MicroMagnetic Framework (OOMMF)~\cite{OOMMF}. The simulation is accomplished by a set of built-in OOMMF extensible solver (OXS) objects. We employ the OXS extension module for modeling the interface-induced antisymmetric exchange interaction, i.e., the Dzyaloshinskii-Moriya interaction (DMI)~\cite{Rohart_PRB2013}. In addition, we use the updated OXS extension module for simulating the in-plane current-induced STTs~\cite{Xichao_CIP2016}. The in-plane current-driven magnetization dynamics is governed by the Landau-Lifshitz-Gilbert (LLG) equation augmented with the adiabatic and non-adiabatic STTs~\cite{LLGCIP,OOMMF}
\begin{align}
\label{LLGS-CIP}
\frac{d\boldsymbol{M}}{dt}=&-\gamma_{0}\boldsymbol{M}\times\boldsymbol{H}_{\text{eff}}+\frac{\alpha}{M_{\text{S}}}(\boldsymbol{M}\times \frac{d\boldsymbol{M}}{dt}) \\ \notag
&+\frac{u}{M_{\text{S}}^2}(\boldsymbol{M}\times\frac{\partial\boldsymbol{M}}{\partial x}\times\boldsymbol{M})-\frac{\beta u}{M_{\text{S}}}(\boldsymbol{M}\times\frac{\partial\boldsymbol{M}}{\partial x}),
\end{align}
where $\boldsymbol{M}$ is the magnetization, $M_{\text{S}}$ is the saturation magnetization, $t$ is the time, $\gamma_{\text{0}}$ is the Gilbert gyromagnetic ratio, $\alpha$ is the Gilbert damping coefficient, and $\beta$ is the strength of the non-adiabatic STT. The adiabatic STT coefficient is given by $u$, i.e., the conduction electron velocity. The effective field $\boldsymbol{H}_{\text{eff}}$ is expressed as
\begin{equation}
\boldsymbol{H}_{\text{eff}}=-\mu_{0}^{-1}\frac{\partial E}{\partial\boldsymbol{M}},
\label{effective-field}
\end{equation}
where $\mu_0$ is the vacuum permeability constant. The average energy density $E$ contains the exchange, anisotropy, demagnetization, and DMI energies, which is given as
\begin{align}
\label{energy-density}
E&=A[\nabla(\frac{\boldsymbol{M}}{M_{\text{S}}})]^{2}-K\frac{(\boldsymbol{n}\cdot\boldsymbol{M})^{2}}{M_{\text{S}}^{2}}-\frac{\mu_{0}}{2}\boldsymbol{M}\cdot\boldsymbol{H}_{\text{d}}(\boldsymbol{M}) \\ \notag
&+\frac{D}{M_{\text{S}}^{2}}(M_{z}\frac{\partial M_{x}}{\partial x}+M_{z}\frac{\partial M_{y}}{\partial y}-M_{x}\frac{\partial M_{z}}{\partial x}-M_{y}\frac{\partial M_{z}}{\partial y}),
\end{align}
where $A$, $K$, and $D$ are the exchange, anisotropy, and DMI energy constants, respectively. $\boldsymbol{n}$ is the unit surface normal vector, and $\boldsymbol{H}_{\text{d}}(\boldsymbol{M})$ is the demagnetization field. $M_x$, $M_y$ and $M_z$ are the three Cartesian components of $\boldsymbol{M}$.

The model is discretized into tetragonal volume elements with the size of $2$ nm $\times$ $2$ nm $\times$ $1$ nm, which ensures a good compromise between the computational accuracy and efficiency. The magnetic parameters are adopted from Refs.~\cite{Sampaio_NNANO2013,Xichao_SREP2015B}:
$\gamma_{0}=2.211\times 10^{5}$ m/(A$\cdot$s),
$A=15$ pJ/m,
$D=3$ mJ/m$^{2}$,
$K=0.8$ MJ/m$^{3}$,
$M_{\text{S}}=580$ kA/m.
In all simulations, we assume $u=100$ m/s and $w=50$ nm. The skyrmion is initially located at the position of $x=100$ nm, $y=25$ nm.

The Gilbert damping coefficient $\alpha$ is defined as a function of the longitudinal coordinate $x$ as follows [Fig.~\ref{FIG1}(a)]
\begin{equation}
\alpha(x)=\alpha_{\text{amp}}\cdot\left\{1+\sin\left[2\pi(x/\lambda_{\alpha})\right]\right\}+\alpha_{\text{min}},
\label{eq:alpha-function}
\end{equation}
where $\alpha_{\text{amp}}=(\alpha_{\text{max}}-\alpha_{\text{min}})/2$ is the amplitude of the $\alpha$ function. $\alpha_{\text{max}}$ and $\alpha_{\text{min}}$ stand for the maximum and minimum values of the $\alpha$ function, respectively. $\lambda_{\alpha}$ denotes the wavelength of the $\alpha$ function.
It is worth mentioning that the spatially varying $\alpha$ can be achieved by gradient doping of lanthanides impurities in ferromagnets~\cite{Reidy_APL2003,He_APL2007,Yuan_AIPADV2015}. Experiments have found that $\alpha$ is dependent on the interface~\cite{Urban_PRL2001}. Thus it is also realistic to construct the varying $\alpha$ by techniques such as interface engineering. Indeed, as shown in Ref.~\cite{King_APL2014}, local control of $\alpha$ in a ferromagnetic/non-magnetic thin-film bilayer has been experimentally demonstrated by interfacial intermixing induced by focused ion-beam irradiation.

In a similar way, the non-adiabatic STT coefficient $\beta$ is also defined as a function of the longitudinal coordinate $x$ as follows [Fig.~\ref{FIG1}(a)]
\begin{equation}
\beta(x)=\beta_{\text{amp}}\cdot\left\{1+\sin\left[2\pi(x/\lambda_{\beta})-\varphi\right]\right\}+\beta_{\text{min}},
\label{eq:beta-function}
\end{equation}
where $\beta_{\text{amp}}=(\beta_{\text{max}}-\beta_{\text{min}})/2$ is the amplitude of the $\beta$ function. $\beta_{\text{max}}$ and $\beta_{\text{min}}$ stand for the maximum and minimum values of the $\beta$ function, respectively. $\lambda_{\beta}$ and $\varphi$ denote the wavelength and phase of the $\beta$ function, respectively.
Since the value of $\beta$ depends on the material properties~\cite{Gilmore_PRB2011}, it is expected to realize the spatial varying $\beta$ by constructing a superlattice nanotrack using different materials, similar to the model given in Ref.~\cite{Yuan_AIPADV2015}. Note that the effect of varying $\beta$ has also been studied in spin torque oscillators~\cite{Zhou_APL2009}.

\begin{figure}[t]
\centerline{\includegraphics[width=0.5\textwidth]{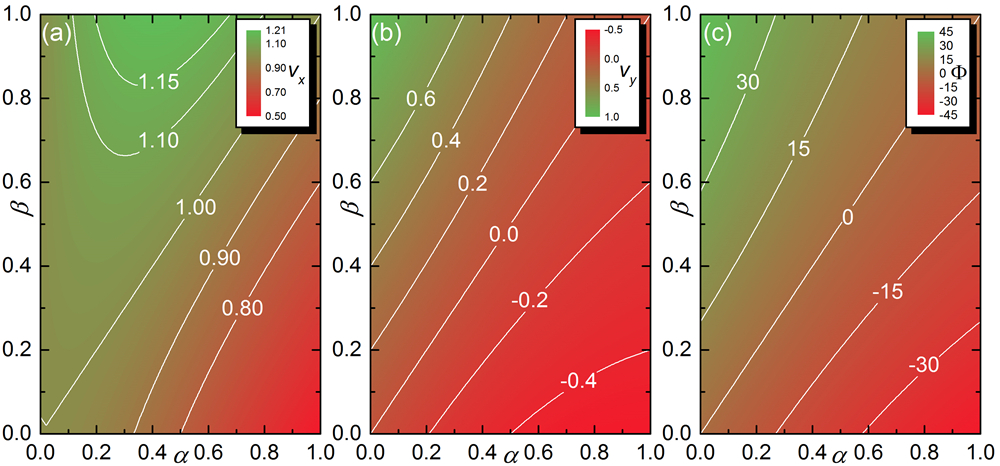}}
\caption{%
(a) $v_x$, (b) $v_y$, and (c) $\Phi$ as functions of $\alpha$ and $\beta$ given by Eq.~(\ref{eq:vx-vy}) and Eq.~(\ref{eq:skyrmion-Hall-angle-Thiele}), respectively. $v_x$ and $v_y$ are reduced by $u$.
}
\label{FIG2}
\end{figure}

\section{Results}
\label{se:Results}

\subsection{Nanotrack with spatially uniform $\alpha$ and $\beta$}
\label{se:uniform-alpha-beta}

We first recapitulate the in-plane current-driven skyrmion motion in a nanotrack with spatially uniform $\alpha$ and $\beta$. As shown in Fig.~\ref{FIG1}(b), the skyrmion moves along the central line of the nanotrack when $\beta=\alpha=0.3$. However, due to the SkHE, it shows a transverse shift toward the upper and lower edges when $\beta=2\alpha=0.6$ and $\beta=\alpha/2=0.15$, respectively. The skyrmion is destroyed by touching the upper edge when $\beta=2\alpha=0.6$ at $t=870$ ps.

The skyrmion Hall angle $\Phi$, which characterizes the transverse motion of the skyrmion caused by the SkHE, is defined as
\begin{equation}
\Phi=\tan^{-1}(v_{y}/v_{x}).
\label{eq:skyrmion-Hall-angle}
\end{equation}
Figure~\ref{FIG1}(c) shows $\Phi$ as a function of $x$ for the skyrmion motion with $\beta=\alpha/2=0.15$, $\beta=\alpha=0.3$, and $\beta=2\alpha=0.6$. It can be seen that $\Phi=0^{\circ}$ when $\beta=\alpha=0.3$, indicating the moving skyrmion has no transverse motion [Fig.~\ref{FIG1}(d)].
When $\beta=\alpha/2=0.15$, $\Phi$ increases from $-15^{\circ}$ to $0^{\circ}$, indicating the moving skyrmion has a transverse shift toward the lower edge which is balanced by the transverse force due to the SkHE and the edge-skyrmion repulsive force [Fig.~\ref{FIG1}(d)].
When $\beta=2\alpha=0.6$, $\Phi$ decreases from $15^{\circ}$ to $3^{\circ}$ within $870$ ps, indicating the moving skyrmion shows a transverse motion toward the upper edge. At $t=870$ ps, the skyrmion is destroyed as it touches the upper edge of the nanotrack [Fig.~\ref{FIG1}(d)]. It should be noted that the skyrmion profile is rigid before it touches the nanotrack edge.
In order to better understand the transverse motion caused by the SkHE, we also analyze the in-plane current-driven skyrmion motion using the Thiele equation~\cite{Thiele_PRL1973,Schulz_NPHYS2012,Everschor_PRB2011,Iwasaki_NCOMMS2013} by assuming the skyrmion moves in an infinite film, which is expressed as
\begin{equation}
\boldsymbol{G}\times\left(\boldsymbol{v}-\boldsymbol{u}\right)+\boldsymbol{\mathcal{D}}\left(\beta\boldsymbol{u}-\alpha\boldsymbol{v}\right)=\boldsymbol{0},
\label{CIP:Thiele-equation}
\end{equation}
where $\boldsymbol{G}=\left(0,0,-4\pi Q\right)$ is the gyromagnetic coupling vector with the skyrmion number
\begin{equation}
Q=\frac{1}{4\pi}\int\boldsymbol{m}\cdot\left(\frac{\partial\boldsymbol{m}}{\partial x}\times\frac{\partial\boldsymbol{m}}{\partial y}\right)dxdy.
\label{eq:Q}
\end{equation}
$\boldsymbol{m}=\boldsymbol{M}/M_{\text{S}}$ is the reduced magnetization and $\boldsymbol{\mathcal{D}}$ is the dissipative tensor
\begin{equation}
\boldsymbol{\mathcal{D}}=4\pi\left(\begin{matrix}\mathcal{D}_{xx} & \mathcal{D}_{xy} \\ \mathcal{D}_{yx} & \mathcal{D}_{yy}\end{matrix}\right).
\label{eq:D}
\end{equation}
$\boldsymbol{u}=(u,0)$ is the conduction electron velocity, and $\boldsymbol{v}$ is the skyrmion velocity. For the nanoscale skyrmion studied here, we have
\begin{equation}
Q=-1, \quad
\mathcal{D}_{xx}=\mathcal{D}_{yy}=1, \quad
\mathcal{D}_{xy}=\mathcal{D}_{yx}=0.
\label{eq:Q-Dxx-Dxy}
\end{equation}
Hence, the skyrmion velocity is given as
\begin{equation}
v_{x}=u\frac{(\alpha\beta+1)}{\alpha^2+1}, \quad
v_{y}=u\frac{(\beta-\alpha)}{\alpha^2+1}.
\label{eq:vx-vy}
\end{equation}
The skyrmion Hall angle $\Phi$ is thus given as
\begin{equation}
\Phi=\tan^{-1}(v_{y}/v_{x})=\tan^{-1}\left(\frac{\beta-\alpha}{\alpha\beta+1}\right).
\label{eq:skyrmion-Hall-angle-Thiele}
\end{equation}
By calculating Eq.~(\ref{eq:vx-vy}), we show $v_x$ as functions of $\alpha$ and $\beta$ in Fig.~\ref{FIG2}(a). $v_x$ ranges between $0.5u$ and $1.21u$, indicating the skyrmion always moves in the $+x$ direction. When $\alpha=0.42$ and $\beta=1$, $v_{x}$ can reach the maximum value of $v_{x}=1.21u$.
Similarly, we show $v_y$ as functions of $\alpha$ and $\beta$ in Fig.~\ref{FIG2}(b). $v_y$ ranges between $-0.5u$ and $u$, indicating the skyrmion can move in both the $\pm y$ directions. When $\alpha<\beta$, $v_{y}>0$, the skyrmion shows a positive transverse motion, while when $\alpha>\beta$, $v_{y}<0$, the skyrmion shows a negative transverse motion.
By calculating Eq.~(\ref{eq:skyrmion-Hall-angle-Thiele}), we also show $\Phi$ as functions of $\alpha$ and $\beta$ in Fig.~\ref{FIG2}(c), where $\Phi$ varies between $\Phi=45^{\circ}$ and $\Phi=-45^{\circ}$. Obviously, one has $\Phi=0^{\circ}$, $\Phi<0^{\circ}$, and $\Phi>0^{\circ}$ for $\alpha=\beta$, $\alpha>\beta$, and $\alpha<\beta$, respectively, which agree with the simulation results for the nanotrack when the edge effect is not significant, i.e., when the skyrmion moves in the interior of the nanotrack. For example, using Eq.~(\ref{eq:skyrmion-Hall-angle-Thiele}), the skyrmion has $\Phi=14^{\circ}$ and $\Phi=-14^{\circ}$ for $\beta=2\alpha=0.6$ and $\beta=\alpha/2=0.15$, respectively, which match the simulation results at $t\sim 0$ ps where the edge effect is negligible [Fig.~\ref{FIG1}(c)].

\begin{figure}[t]
\centerline{\includegraphics[width=0.5\textwidth]{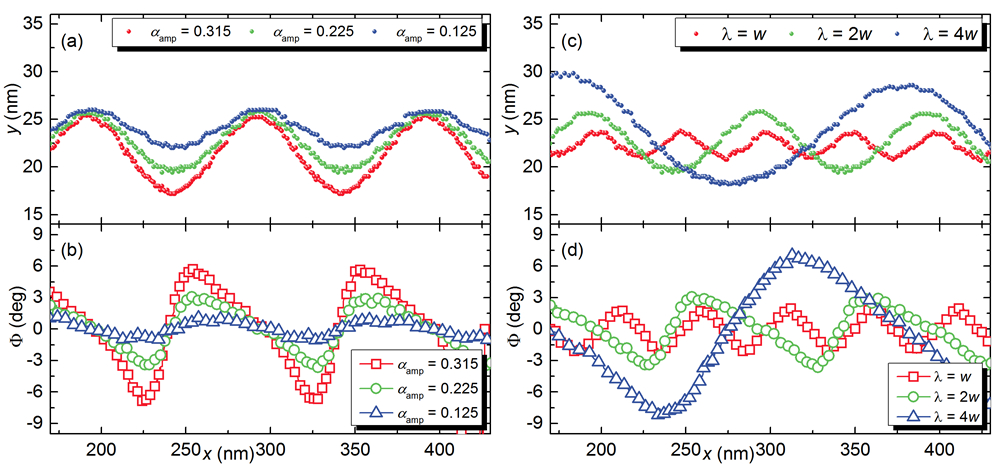}}
\caption{%
(a) Trajectories of current-driven skyrmions with $\alpha_{\text{amp}}=0.315,0.225,0.215$. $\lambda_{\alpha}=2w$ and $\beta=0.3$.
(b) $\Phi$ as a function of $x$ for skyrmion motion with $\alpha_{\text{amp}}=0.315,0.225,0.215$. $\lambda_{\alpha}=2w$ and $\beta=0.3$.
(c) Trajectories of current-driven skyrmions with $\lambda_{\alpha}=w,2w,4w$. $\alpha_{\text{amp}}=0.225$ and $\beta=0.3$.
(d) $\Phi$ as a function of $x$ for skyrmion motion with $\lambda_{\alpha}=w,2w,4w$. $\alpha_{\text{amp}}=0.225$ and $\beta=0.3$.
}
\label{FIG3}
\end{figure}

\begin{figure}[t]
\centerline{\includegraphics[width=0.5\textwidth]{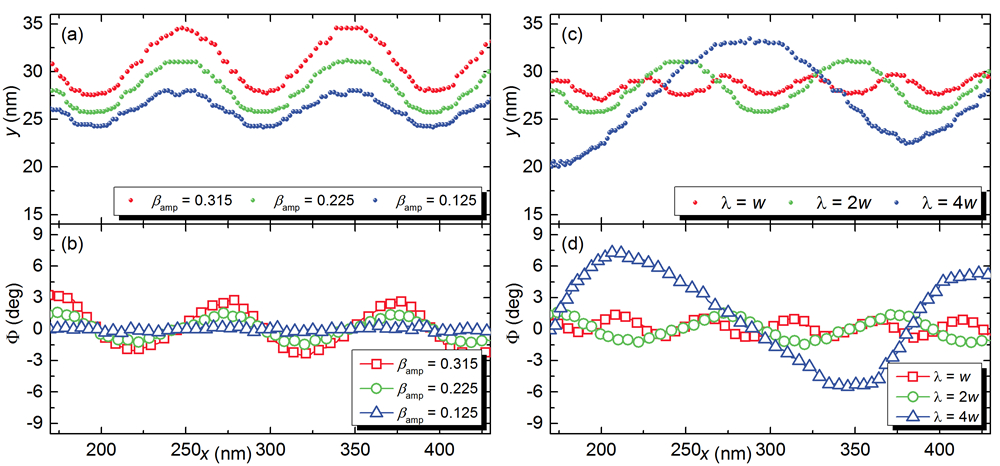}}
\caption{%
(a) Trajectories of current-driven skyrmions with $\beta_{\text{amp}}=0.315,0.225,0.215$. $\lambda_{\beta}=2w$, $\varphi=0$, and $\alpha=0.3$.
(b) $\Phi$ as a function of $x$ for skyrmion motion with $\beta_{\text{amp}}=0.315,0.225,0.215$. $\lambda_{\beta}=2w$, $\varphi=0$, and $\alpha=0.3$.
(c) Trajectories of current-driven skyrmions with $\lambda_{\beta}=w,2w,4w$. $\beta_{\text{amp}}=0.225$, $\varphi=0$, and $\alpha=0.3$.
(d) $\Phi$ as a function of $x$ for skyrmion motion with $\lambda_{\beta}=w,2w,4w$. $\beta_{\text{amp}}=0.225$, $\varphi=0$, and $\alpha=0.3$.
}
\label{FIG4}
\end{figure}

\subsection{Nanotrack with spatially varying $\alpha$ or $\beta$}
\label{se:varying-alpha-or-varying-beta}

We first demonstrate the in-plane current-driven skyrmion motion in a nanotrack with spatially varying $\alpha$ and spatially uniform $\beta$, i.e., $\alpha$ is a function of $x$, as in Eq.~(\ref{eq:alpha-function}), and $\beta=0.3$.
Figure~\ref{FIG3}(a) shows the trajectories of the current-driven skyrmions with different $\alpha(x)$ functions where $\lambda_{\alpha}=2w$ and $\beta=0.3$. For $\alpha_{\text{max}}=0.75$, $\alpha_{\text{min}}=0.12$, i.e., $\alpha_{\text{amp}}=0.315$, the skyrmion moves in the rightward direction in a sinusoidal pattern. For $\alpha_{\text{max}}=0.6$, $\alpha_{\text{min}}=0.15$, i.e., $\alpha_{\text{amp}}=0.225$, the maximum transverse shift of skyrmion is reduced in compared to that of $\alpha_{\text{amp}}=0.315$. For $\alpha_{\text{max}}=0.45$, $\alpha_{\text{min}}=0.2$, i.e., $\alpha_{\text{amp}}=0.125$, the amplitude of the skyrmion trajectory further decreases. $\Phi$ as a function of $x$ corresponding to Fig.~\ref{FIG3}(a) for different $\alpha(x)$ functions are given in Fig.~\ref{FIG3}(b).
Figure~\ref{FIG3}(c) shows the trajectories of the current-driven skyrmions with different $\lambda_{\alpha}$ where $\alpha_{\text{amp}}=0.225$ and $\beta=0.3$. $\Phi$ as a function of $x$ corresponding to Fig.~\ref{FIG3}(c) for different $\lambda_{\alpha}$ are given in Fig.~\ref{FIG3}(d).

We then investigate the in-plane current-driven skyrmion motion in a nanotrack with spatially uniform $\alpha$ and spatially varying $\beta$, i.e., $\beta$ is a function of $x$, as in Eq.~(\ref{eq:beta-function}), and $\alpha=0.3$.
Figure~\ref{FIG4}(a) shows the trajectories of the current-driven skyrmions with different $\beta(x)$ functions where $\lambda_{\beta}=2w$, $\varphi=0$ and $\alpha=0.3$. The results are similar to the case with spatially varying $\alpha$. For $\beta_{\text{max}}=0.75$, $\beta_{\text{min}}=0.12$, i.e., $\beta_{\text{amp}}=0.315$, the skyrmion moves in the rightward direction in a sinusoidal pattern. For $\beta_{\text{max}}=0.6$, $\beta_{\text{min}}=0.15$, i.e., $\beta_{\text{amp}}=0.225$, the maximum transverse shift of skyrmion is reduced in compared to that of $\beta_{\text{amp}}=0.315$. For $\beta_{\text{max}}=0.45$, $\beta_{\text{min}}=0.2$, i.e., $\beta_{\text{amp}}=0.125$, the amplitude of the skyrmion trajectory further decreases. $\Phi$ as a function of $x$ corresponding to Fig.~\ref{FIG4}(a) for different $\beta(x)$ functions are given in Fig.~\ref{FIG4}(b).
Figure~\ref{FIG4}(c) shows the trajectories of the current-driven skyrmions with different $\lambda_{\beta}$ where $\beta_{\text{amp}}=0.225$ and $\alpha=0.3$. $\Phi$ as a function of $x$ corresponding to Fig.~\ref{FIG4}(c) for different $\lambda_{\beta}$ are given in Fig.~\ref{FIG4}(d).

\begin{figure}[t]
\centerline{\includegraphics[width=0.5\textwidth]{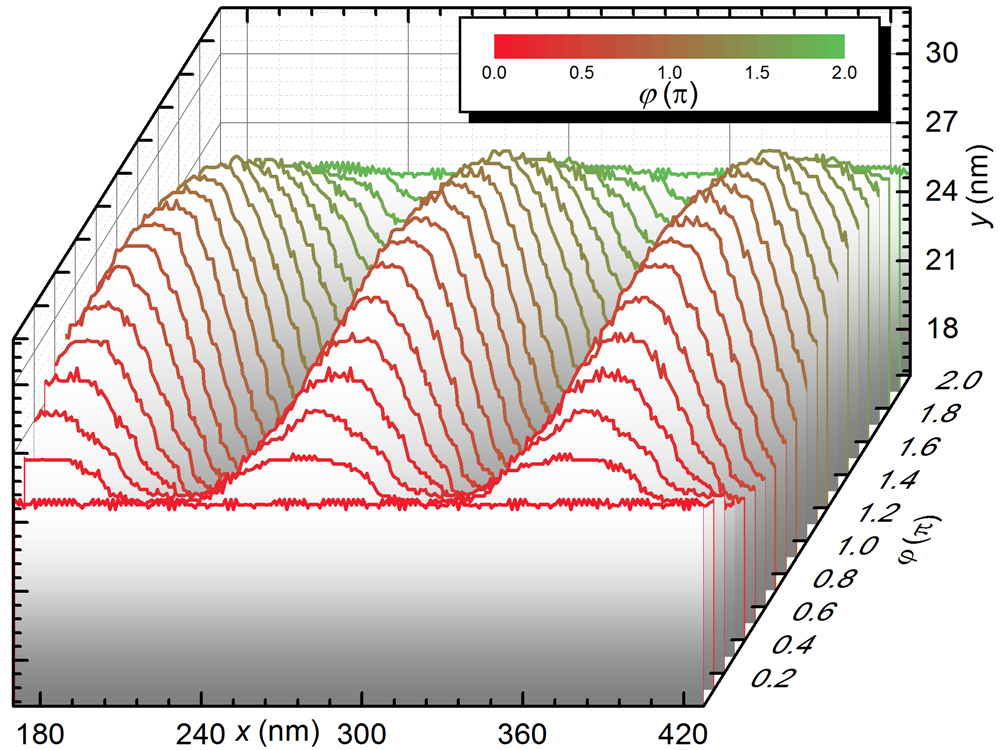}}
\caption{%
Trajectories of current-driven skyrmions with $\varphi=0\sim 2\pi$. $\alpha_{\text{amp}}=\beta_{\text{amp}}=0.225$ and $\lambda_{\alpha}=\lambda_{\beta}=2w$.
}
\label{FIG5}
\end{figure}

From the skyrmion motion with spatially varying $\alpha$ or spatially varying $\beta$, it can be seen that the amplitude of trajectory is proportional to $\alpha_{\text{amp}}$ or $\beta_{\text{amp}}$. The wavelength of trajectory is equal to $\lambda_{\alpha,\beta}$, while the amplitude of trajectory is proportional to $\lambda_{\alpha,\beta}$. $\Phi$ also varies with $x$ in a quasi-sinusoidal manner, where the peak value of $\Phi(x)$ is proportional to $\alpha_{\text{amp}}$, $\beta_{\text{amp}}$, and $\lambda_{\alpha,\beta}$. As shown in Fig.~\ref{FIG2}(c), when $\beta$ is fixed at a value between $\alpha_{\text{max}}$ and $\alpha_{\text{min}}$, larger $\alpha_{\text{amp}}$ will lead to larger peak value of $\Phi(x)$. On the other hand, a larger $\lambda_{\alpha,\beta}$ allows a longer time for the skyrmion transverse motion toward a certain direction, which will result in a larger amplitude of trajectory as well as a larger peak value of $\Phi(x)$.

\begin{figure}[t]
\centerline{\includegraphics[width=0.5\textwidth]{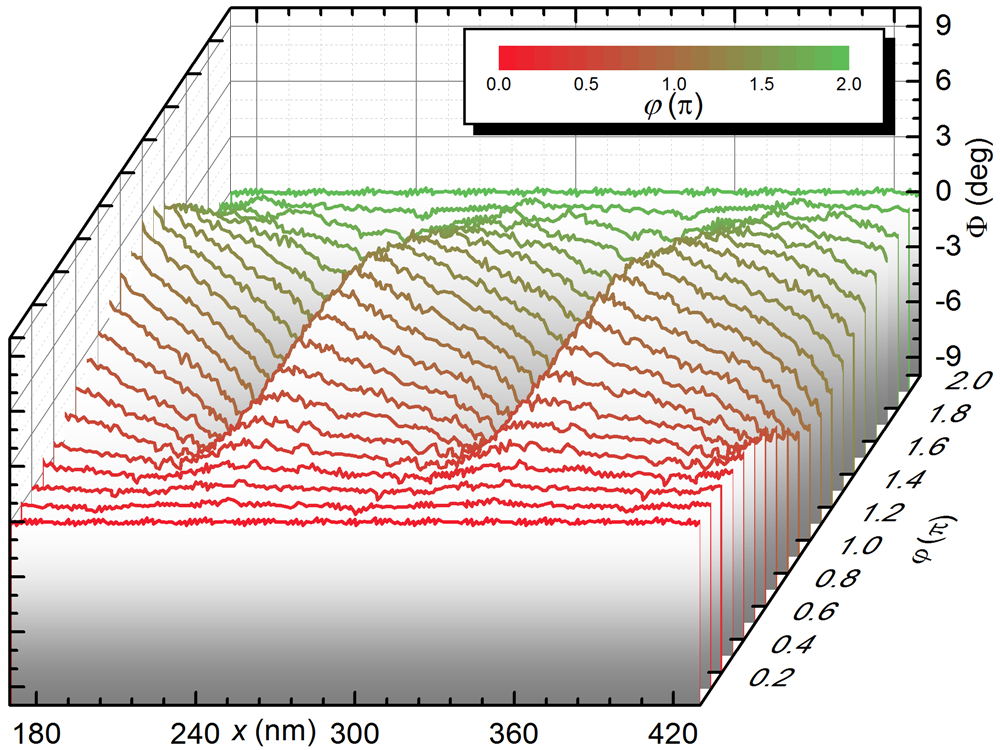}}
\caption{%
$\Phi$ as a function of $x$ for skyrmion motion with $\varphi=0\sim 2\pi$. $\alpha_{\text{amp}}=\beta_{\text{amp}}=0.225$ and $\lambda_{\alpha}=\lambda_{\beta}=2w$.
}
\label{FIG6}
\end{figure}

\subsection{Nanotrack with spatially varying $\alpha$ and $\beta$}
\label{se:varying-alpha-varying-beta}

We also demonstrate the in-plane current-driven skyrmion motion in a nanotrack with both spatially varying $\alpha$ and $\beta$, i.e., both $\alpha$ and $\beta$ are functions of $x$, as given in Eq.~(\ref{eq:alpha-function}) and Eq.~(\ref{eq:beta-function}), respectively.

Figure~\ref{FIG5} shows the trajectories of the current-driven skyrmions with spatially varying $\alpha$ and $\beta$ where $\alpha_{\text{amp}}=\beta_{\text{amp}}=0.225$ and $\lambda_{\alpha}=\lambda_{\beta}=2w$. Here, we focus on the effect of the phase difference between the $\alpha(x)$ and $\beta(x)$ functions. For $\varphi=0$ and $\varphi=2\pi$, as the $\alpha(x)$ function is identical to the $\beta(x)$ function, the skyrmion moves along the central line of the nanotrack. For $0<\varphi<2\pi$, as $\alpha(x)$ could be different from $\beta(x)$ at a certain $x$, it is shown that the skyrmion moves toward the right direction in a sinusoidal pattern, where the phase of trajectory is subject to $\varphi$.
Figure~\ref{FIG6} shows $\Phi$ as a function of $x$ corresponding to Fig.~\ref{FIG5} for $\varphi=0\sim 2\pi$ where $\alpha_{\text{amp}}=\beta_{\text{amp}}=0.225$ and $\lambda_{\alpha}=\lambda_{\beta}=2w$. It shows that $\Phi=0^{\circ}$ when $\varphi=0$ and $\varphi=2\pi$, while it varies with $x$ in a quasi-sinusoidal manner when $0<\varphi<2\pi$. The amplitude of trajectory as well as the peak value of $\Phi(x)$ reach their maximum values when $\varphi=\pi$.

\section{Conclusion}
\label{se:Conclusion}

In conclusion, we have shown the in-plane current-driven motion of a skyrmion in a nanotrack with spatially uniform $\alpha$ and $\beta$, where $\Phi$ is determined by $\alpha$ and $\beta$, which can vary between $\Phi=45^{\circ}$ and $\Phi=-45^{\circ}$ in principle.
Then, we have investigated the in-plane current-driven skyrmion motion in a nanotrack with spatially sinusoidally varying $\alpha$ or $\beta$. The skyrmion moves on a sinusoidal trajectory, where the amplitude and wavelength of trajectory can be controlled by the spatial profiles of $\alpha$ and $\beta$. The peak value of $\Phi(x)$ is proportional to the amplitudes and wavelengths of $\alpha(x)$ and $\beta(x)$.
In addition, we have demonstrated the in-plane current-driven skyrmion motion in a nanotrack having both spatially sinusoidally varying $\alpha$ and $\beta$ with the same amplitude and wavelength. The skyrmion moves straight along the central line of the nanotrack when $\alpha(x)$ and $\beta(x)$ have no phase difference, i.e., $\varphi=0$. When $\varphi\neq 0$, the skyrmion moves in a sinusoidal pattern, where the peak value of $\Phi(x)$ reaches its maximum value when $\varphi=\pi$.
This work points out the possibility to guide and control skyrmion motion in a nanotrack by constructing spatially varying parameters, where the destruction of skyrmion caused by the SkHE can be prevented, which enables reliable skyrmion transport in skyrmion-based information processing devices.

\section*{Acknowledgment}

X.Z. was supported by JSPS RONPAKU (Dissertation Ph.D.) Program.
G.P.Z. was supported by the National Natural Science Foundation of China (Grants No. 11074179 and No. 10747007), and the Construction Plan for Scientific Research Innovation Teams of Universities in Sichuan (No. 12TD008).
Y.Z. was supported by the Shenzhen Fundamental Research Fund under Grant No. JCYJ20160331164412545.



\end{document}